\documentclass{article}
\usepackage{spconf,amsmath,graphicx}
\usepackage{hyperref}
\usepackage{url}
\usepackage{booktabs}
\usepackage{bm}
\usepackage{array}
\usepackage{cite}
\usepackage{multirow}
\usepackage{units}
\usepackage[acronym]{glossaries}
\usepackage{enumitem}

\newacronym{ASR}{ASR}{Automatic Speech Recognition}
\newacronym{WER}{WER}{Word Error Rate}
\newacronym{SNR}{SNR}{Signal-to-Noise Ratio}
\newacronym{MOS}{MOS}{Mean Opinion Score}

\usepackage[skip=10pt]{caption}

\def\s{{\mathbf s}}
\def\L{{\cal L}}
\def\mv#1{\boldsymbol{{#1}}}
\def\mvt#1{\mv{\tilde{#1}}}
\def\mc#1{{\mathcal{#1}}}

\def\mv#1{\boldsymbol{{#1}}}
\def\mvt#1{\mv{\tilde{#1}}}
\def\mc#1{{\mathcal{#1}}}

\def\stilde{\bm{\tilde{s}}}

\def\Stilde{\bm{\tilde{S}}}

\def\L{\mathcal{L}}

\def\L{{\cal L}}

\def\L{{\cal L}}

\title{Egocentric Audio-Visual Noise Suppression}
\name{Roshan Sharma$^1$\sthanks{The first author performed the work while at Meta },Weipeng He$^2$, Ju Lin$^2$, Egor Lakomkin$^2$, Yang Liu$^2$, Kaustubh Kalgaonkar$^2$}
\address{$^1$Carnegie Mellon University, Pittsburgh, USA \\ $^2$Meta, Seattle, USA
}
\begin{document}
\ninept
\maketitle
\begin{abstract}

This paper studies audio-visual noise suppression for egocentric videos -- where the speaker is not captured in the video. Instead, potential noise sources are visible on screen with the camera emulating the off-screen speaker's view of the outside world. This setting is different from prior work in audio-visual speech enhancement that relies on lip and facial visuals.
In this paper, we first demonstrate that egocentric visual information is helpful for noise suppression. We compare object recognition and action classification-based visual feature extractors and investigate methods to align audio and visual representations. Then, we examine different fusion strategies for the aligned features, and locations within the noise suppression model to incorporate visual information. Experiments demonstrate that visual features are most helpful when used to generate additive correction masks. Finally, in order to ensure that the visual features are discriminative with respect to different noise types, we introduce a multi-task learning framework that jointly optimizes audio-visual noise suppression and video-based acoustic event detection. This proposed multi-task framework outperforms the audio-only baseline on all metrics, including a 0.16 PESQ improvement. Extensive ablations reveal the improved performance of the proposed model with multiple active distractors, overall noise types, and across different SNRs.  
\end{abstract}
\section{Introduction}
\label{sec:intro}



Speech extracted from the wild is rich in information content. However, such speech often contains noise, which decreases the its intelligibility. Noise suppression is the task of learning to produce cleaner output speech from noisy input. 

Noise suppression can benefit from additional information in the form of visual representations~\cite{10.1007/3-540-40063-X_5,partan1999communication} of the speaker and their environment ~\cite{sharma2022} Prior work in audio-visual noise suppression has examined visual information with on-screen speakers derived from videos~\cite{afouras2018conversation} and still images~\cite{chung20c_interspeech} of the speaker. These visuals capture the motion of the speaker's lips and other articulators, providing information on what is spoken. Such visual information has also been used for related tasks like multi-source audio-visual separation~\cite{Audioscope21,tzinis2022audioscopev2}, object detection~\cite{Afouras_2022_CVPR}, and source localization~\cite{Zhu_2022_WACV}. However, such visuals (with the target speaker on-screen) are challenging to capture in the wild. 

Another source of visual information that could assist noise suppression is visual cues from the surrounding environment. Such a view of the world through the speaker's eyes is termed "egocentric". Egocentric visuals capture surrounding objects that may produce noise, but not the speaker's visual attributes, i.e., the speaker is off-screen. To the best of our knowledge, this setting is vastly different from previous work, which have relied on visual representations of the on-screen speaker for noise suppression. In order to use egocentric visuals for noise suppression, we propose to extract visual features that describe potential distractor sources on screen as it is known that an explicit characterization of the noise source aids noise suppression performance~\cite{boll1979suppression}.

Visual features that represent potential distractor sources may be obtained using object detection models which provide information on the stationary objects, or using video classification models which use longer temporal context to identify on-screen actions. However, such representations may not sufficiently represent the distractor sources. When multiple objects or events are visible on screen, some or none of them may correspond to acoustic events and noise sources.  This motivates us to provide additional supervision over the visual features to make them relevant to the acoustic events in the scene. Therefore, the problem of egocentric audio-visual noise suppression is formulated using Multi-Task Learning (MTL) to jointly optimize visual acoustic event detection and noise suppression.

Audio and visual features have different rates of information change, and are hence unaligned. In order to align the modalities, we compare two strategies- upsampling and temporal attention. Once the modalities have been aligned, the visual features can be incorporated at different locations within the noise-suppression model; and based on this location, we investigate \textit{input}, \textit{intermediate}, \textit{late}, and \textit{mask fusion}. Finally, in order to fuse the audio and visual representations - we compare addition and concatenation.


In summary, this paper makes the following contributions:
\begin{enumerate}[topsep=0pt,itemsep=-1ex,partopsep=1ex,parsep=1ex]
    \item We propose a method to utilize egocentric visual information to enhance off-screen target speech, and compare visual representations from pretrained models that distinguish potential distractor sources.
    \item In order to produce visual representations that can distinguish between different distractors, we introduce a novel vision based acoustic event detection criterion, which produces better enhancement than pre-trained object or action classifiers. 
    \item We compare temporal attention and upsampling to align the audio and visual feature sequences, and explore addition and concatenation methods to fuse the aligned representations within multiple locations within the model, corresponding to \textit{input}, \textit{intermediate}, \textit{late} and \textit{mask fusion} strategies. 
    
\end{enumerate}





\section{Proposed Approach}
\label{sec:proposed}
\begin{figure}[!ht]
    \includegraphics[width=0.5\textwidth]{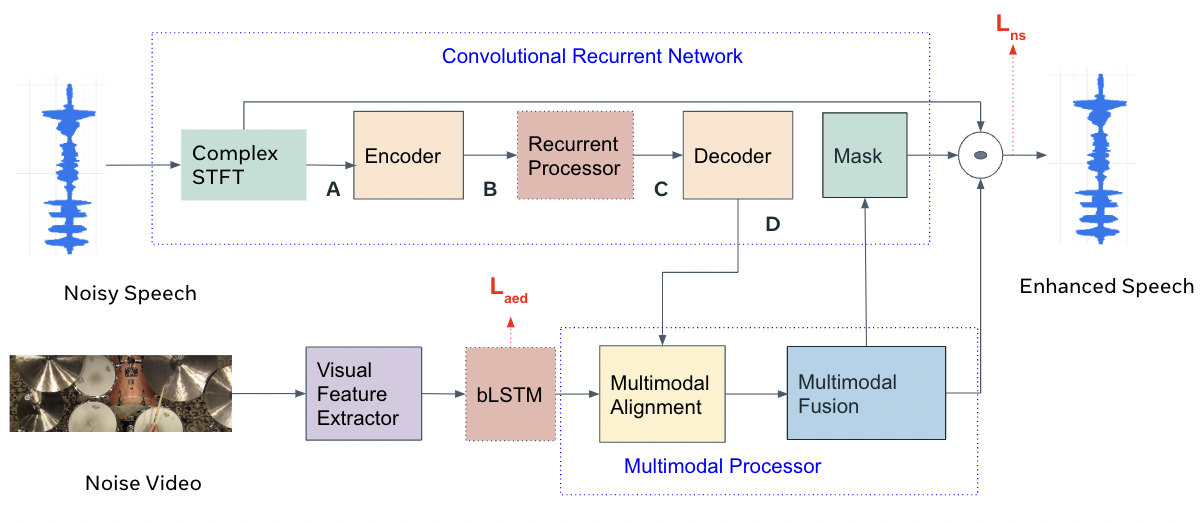}
    \caption{Model architecture of the multi-task audio-visual convolutional recurrent network (CRN) which performs mask based fusion. Markers A, B, C, and D (in black) indicate locations for audio-visual fusion corresponding to input, intermediate, late and output fusion. }
    \label{fig:avferraris_arch}
\end{figure}
Fig.~\ref{fig:avferraris_arch} shows the proposed model architecture for audio-visual noise suppression. The complex valued STFT of the noisy audio input is passed through the Convolutional Recurrent Network (CRN)\cite{tan2018convolutional} to obtain the complex STFT mask, which is then multiplied with the input to generate the predicted clean output. 

The raw video is passed through a visual feature extractor that either extracts per-frame or per-video visual representations. The video features thus obtained are passed through recurrent layers to utilize temporal dependencies. Audio feature maps from the four marked locations (A), (B), (C), or (D) can be used for multimodal alignment and fusion, and the output of the fusion module is then used as input to the next CRN layer.  The multimodal alignment module then either performs deterministic temporal upsampling or uses attention to align the audio and visual feature sequences. Finally the multimodal fusion block uses addition or concatenation to combine the audio and visual feature maps.

\subsection{Convolutional Recurrent Network (CRN)}

The CRN model comprises an encoder, a recurrent processor, and a decoder, and has a U-Net structure that enables feature sharing between the encoder and decoder at various resolutions for efficient optimization.  The encoder consists of 5 encoder blocks- each of which includes a 2-D convolution layer, a 2-D batch normalization layer ~\cite{ioffe2015batch}, and the Gated Linear Unit (GLU)~\cite{dauphin2017language} activation. The convolution layers have kernel sizes of $(2, 4)$ and strides of $(1, 2)$. The first convolution maps the 2 input channels representing the real and imaginary parts of the input spectrogram to 16 output channels, and the subsequent convolutions produce 32, 64, 76, and 98 output channels respectively. The recurrent processor has four bidirectional Long Short-Term Memory (bLSTM) layers with input and hidden
size 294 to capture long-range dependencies. The output is downsized to 
 294, and reshaped to 4 dimensions before being processed by the decoder.

The decoder consists of 5 decoder blocks - each comprising concatenation layers that fuse the encoder outputs, batch normalization layers, and gated transpose convolution layers. The concatenation layer takes as input the decoder activation for block $l$ and the corresponding output activation for the $5-l$ encoder block and stacks them along the channel dimension to form an input with twice the number of input channels. The transpose convolution layers have the same kernel size and stride as the convolution layers in the encoder and produce outputs with 76, 64, 32, 16 and 2 channels respectively. The final output of the decoder with 2 channels is passed through a sigmoid mask layer.  


\subsection{Visual Feature Extractor}
Visual features that represent objects and actions within the video are likely to represent potential noise sources. In this paper, we compare object recognition based visual features with video action classification based features for audio-visual noise suppression.  Object recognition based features derived from \texttt{Efficientnet-B0}\cite{tan2019efficientnet} model pre-trained on Imagenet\cite{deng2009imagenet} to represent the distractor sources. Further, video classification features are useful indicators of actions such as "door open/close" and "playing the guitar" that correspond to noise sources. Action recognition models trained on KINETICS-400\cite{zisserman2017kinetics} are used to extract such representations with pre-trained \texttt{R2+1D}\cite{du2018r2plus1d} model.

\subsection{Multimodal Processor}
\noindent \textbf{Multimodal Alignment}:The audio and visual feature maps have different temporal resolutions as the audio information changes faster than the video. Therefore, it is necessary to obtain a temporal alignment between audio and video. In this paper, we compare upsampling and temporal attention for multimodal alignment. Upsampling is a deterministic mechanism that is favoured when the audio and video sequences are monotonic with respect to each other, i.e., when the first video frame corresponds to the first few audio frames and so on. Multi-head temporal attention is learned and can be used to align the video and audio sequences even when the montonicity criterion is not satisfied. 

\noindent \textbf{Multimodal Fusion}: Given audio and video feature maps with the same temporal resolution, in this paper we compare different locations within the model to determine the optimal location for audio-visual fusion. Within the audio-visual model architecture shown in Fig.~\ref{fig:avferraris_arch}, letters in black demonstrate potential locations for the visual features. They can be placed in location (A) for \textit{input fusion}, (B) for \textit{intermediate fusion}, (C) for \textit{late fusion}, or (D) for \textit{mask fusion}. 

Based on the location of fusion, the audio and visual feature maps are extracted and then combined.  In this paper, two methods are considered for fusion- \textit{addition} and \textit{concatenation}. First, the aligned visual feature map is projected such that it has the same feature dimension as the audio feature map, and one single channel. If addition is the fusion method, both the visual feature map is passed through a convolution layer to have the same number of channels as the audio feature map. Then the resulting maps are added to produce the output. For concatenative fusion, the visual feature map with a single channel is concatenated with the audio feature map along the channel dimension, and the resulting map is transformed to the shape of the input audio feature map using a 2D convolution.  

\subsection{Optimization Criterion}
Our noise suppression models are trained using a combination of several losses. Let $\stilde$ be the enhanced 16 kHz signal in the time domain. Let $\mv{S}$ denote the STFT magnitude of clean speech $\s$. The reconstruction loss includes an $L_1$ loss in the time domain, a weighted STFT loss (W-STFT)~\cite{lin2022speech}, and Scale Invariant Signal Distortion Ratio (SI-SDR) loss~\cite{le2019sdr}.
The total loss $L_{ns}$ is computed as a weighted sum of these components in Eq.~(\ref{eq:ns_loss}), where weights $\lambda_1=1,\lambda_2=22.62$ and $\lambda_3=0.001$ are set empirically. 
\begin{equation}
\L_\text{ns} = \lambda_1||\s - \stilde||_1 + \lambda_2\L_\text{WSTFT}(\mv{S}, \Stilde) + \lambda_3\L_\text{SI-SDR}(\mv{s}, \stilde) .
\label{eq:ns_loss}
\end{equation}

Eq.~(\ref{eq:MSTFT}) formulates the weighted STFT loss, where higher weights are used to emphasize the high frequency regions. We split the frequency bins into 4 sub-bands and assign weights (0.1, 1.0, 1.5, 1.5) to each sub-band empirically.
\begin{equation}
\L_\text{WSTFT} = \sum_{k=1}^{4}w_k||\mv{S}_k - \Stilde_k||_1.
\label{eq:MSTFT}
\end{equation}

SI-SDR is commonly used as a loss function in the time-domain. $\mc{L}_{\text{SI-SDR}}$ is defined as:
\begin{equation} 
\begin{split}
\mc{L}_{\text{SI-SDR}} = 10 \log_{10} \frac{||\alpha\mv{s }||^2}{||\alpha\mv{s}-\mv{\tilde{s}}||^2}\\
\text{where } \alpha = \underset{\alpha}\arg\min||\alpha\mv{s} - \mvt{s}||^2 . \\
\end{split}
\label{eq:sisdr}
\end{equation}


\subsection{Multi-Task Learning: Supervision over Visual Features}

Though pre-trained object and action features capture useful visual information about the distractor sources, they generate noisy representations when there are multiple objects and actions in the scene, some of which may not be linked to the audio event. Therefore, it would be helpful to obtain discriminative visual features that recognize different acoustic events. We propose to optimize the extracted visual features by using visual Acoustic Event Detection (AED) as an auxiliary criterion over the outputs of the bLSTM in Fig.~\ref{fig:avferraris_arch} during training.  A linear projection is used to map the pooled features to the label dimension, and binary cross-entropy loss is computed over each of the labels.

Eq.~(\ref{eq:aed_mtl}) shows the proposed Multi-Task Learning (MTL) framework that optimizes audio-visual noise suppression with loss $L_{ns}$(see Eq.~(\ref{eq:ns_loss})) and a  multilabel binary cross entropy loss $L_{aed}$ for visual acoustic event classification. The task weights $\alpha_1=1,\alpha_2=50$ are set empirically.
\begin{equation}
\begin{split}
L_{total} = \alpha_1 L_{ns} + \alpha_2 L_{aed}. \\    
\end{split}
\label{eq:aed_mtl}
\end{equation}



\section{Experiments}
\label{sec:results}

\begin{table} [tbp]
\caption{Speech Enhancement evaluation metrics for audio and audio-visual models on the Audioset evaluation set.}
\label{tab:ns_eval_audioset}
\resizebox{1.02\columnwidth}{!}{  
\begin{tabular}{l|r|r|r|r}
\hline
\textbf{Model}    & \multicolumn{1}{l|}{\textbf{PESQ}} & \multicolumn{1}{l|}{\textbf{STOI}} & \multicolumn{1}{l|}{\textbf{VISQOL-A}} & \multicolumn{1}{l}{\textbf{VISQOL-S}} \\ \hline
Noisy Speech             & 1.2                             & 0.75                              & 2.66                                   & 1.97                                 \\ \hline 
Audio-only            & 2.44                              & 0.89                              & 3.96                                   & 2.97                                 \\ \hline 
+ Visual (Obj)     & 2.56                              & 0.92                              & 4.08                                   & 3.23                                  \\ 
+ MTL & \textbf{2.60}              & \textbf{0.92}           & \textbf{4.45}                  & \textbf{2.95} \\   \hline
+ Visual (Vid)     & 2.55                             & 0.921                             & 4.42                                    & 2.88                                 \\
+ MTL     & 2.57                             & 0.923                             & 4.45                                   & 2.95                                 \\ \hline 
\end{tabular}

}
\end{table}

\begin{table} [tbp]
\caption{PESQ scores of the proposed audio and audio-visual models across different evaluation SNRs.}
\label{tab:snr_analysis_pesq}
\resizebox{1.02\columnwidth}{!}{  
\begin{tabular}{l|r|r|r|r|l}
\hline
\textbf{Model}     & \multicolumn{1}{l|}{\textbf{- 20 dB}} & \multicolumn{1}{l|}{\textbf{-10 dB}} & \multicolumn{1}{l|}{\textbf{0 dB}} & \multicolumn{1}{l|}{\textbf{10 dB}} & \textbf{20 dB}           \\ \hline
Audio              & 1.31                                  & 1.67                                 & 2.44                               & 3.19                               & 3.68 \\
AV (Obj) & 1.33                                  & 1.75                                 & 2.55                               & 3.28                                & 3.75 \\
+ MTL  & 1.37              & 1.78              & 2.60              &     3.32 & 3.78 \\ \hline       
\end{tabular}
}
\end{table}
\vspace{-3mm}
\begin{figure} [tp]
    \centering
    \includegraphics[width=0.4\textwidth]{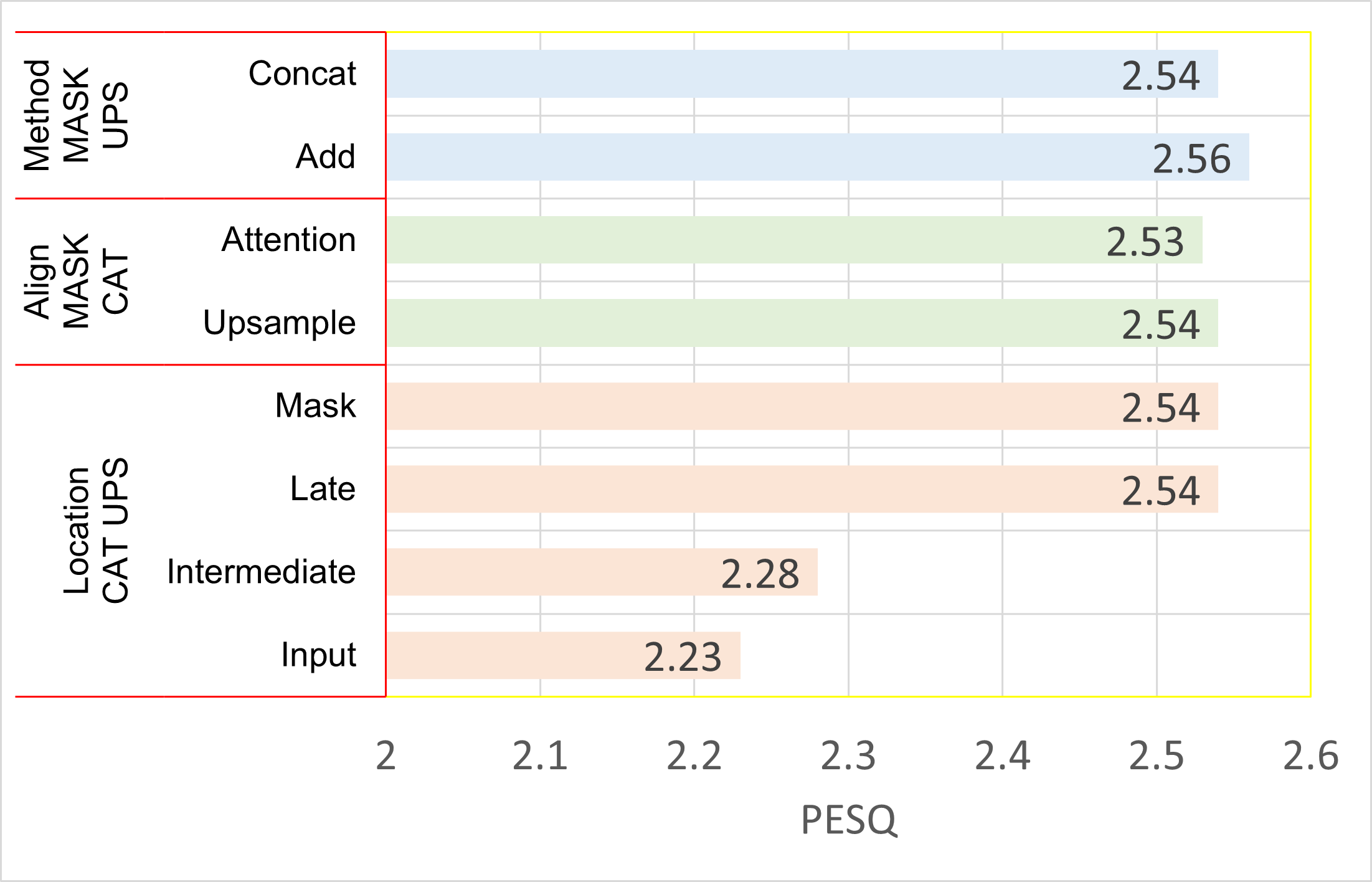}
        \caption{PESQ improvements for different fusion methods(concat, add), alignment methods (upsampling, attention), and locations (input fusion, intermediate fusion, late fusion and mask fusion)}
    \label{fig:pesq_fusion}
\end{figure}

\begin{figure} [t]
    \centering
    \includegraphics[width=0.4\textwidth]{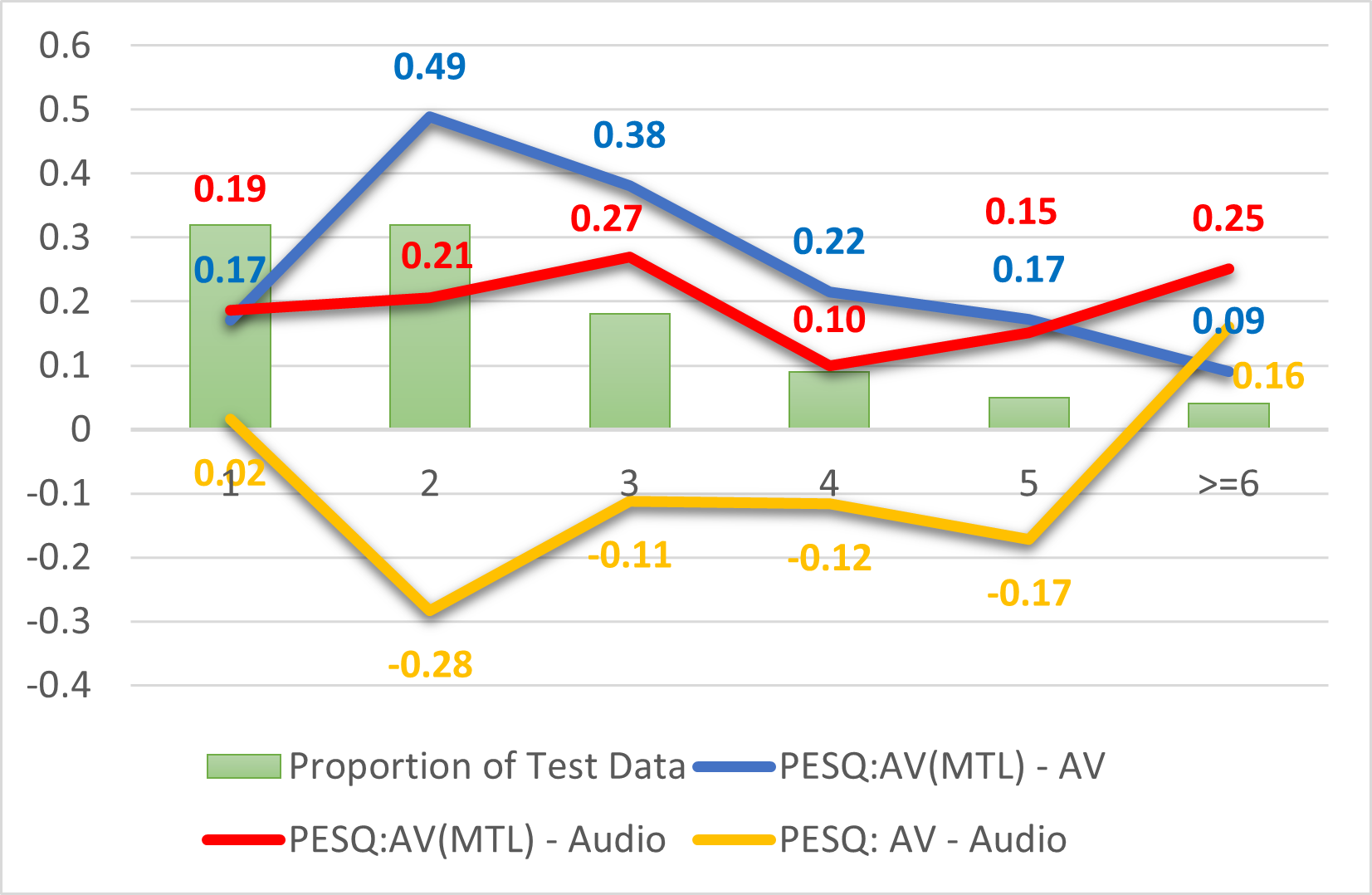}
        \caption{PESQ improvement across the Audio, Audio-visual and Multi-task Audio-visual models for different number of acoustic event labels on the test data. The proportion of the number of sources with a given number of labels is shown in green. The relative improvements in PESQ between Audio and Audio-visual (yellow), Audio-Visual (MTL) and Audio (red), and Audio-visual (MTL) and Audio-visual are also marked plotted.}
    \label{fig:pesq_number_labels}
\end{figure}

\subsection{Dataset and Input Features}

Audioset~\cite{Audioset} is a large scale dataset of videos extracted from YouTube which are human labelled for acoustic event detection. We select portions of the Audioset data that likely do not contain speech as a distractor for our experiments based on manual inspection of the acoustic event labels. We generate data for our experiments by using audio-visual distractor signals from Audioset, and clean off-screen target speech from the DNS Challenge data~\cite{reddy2021interspeech}. The DNS-challenge training data is randomly partitioned into training and evaluation data. The distractor audio and target clean off-screen speech are combined after volume normalization by sampling a mixing Signal-to-Noise Ratio (SNR) $s \sim N(0,5)$.  The test set is created using the same mechanism by mixing evaluation set videos from Audioset with our evaluation partition of the DNS challenge training data.

\subsection{Evaluation Metrics}

Our models are evaluated using both objective and subjective speech quality metrics. For objective evaluation, we use Perceptual Evaluation of Speech Quality (PESQ)\footnote{https://pypi.org/project/pesq/}~\cite{PESQ}, Short-Time Objective Inteligibility (STOI)~\cite{STOI}, and Virtual Speech Quality Objective Listener (VISQOL)~\cite{VISQOL}. We also obtain Deep Noise Suppression Mean Opinion Score (DNSMOS) ~\cite{reddy2021dnsmos} for the best model as a proxy for subjective human evaluation.

\subsection{Audio-Visual Models}
 The audio only CRN model is trained first and used to initialize the parameters of the audio-visual model. This ensures that the audio-visual model does not learn to rely exclusively on the visual representations, leading to noisy predictions. The visual feature extractors are first frozen, and then fine-tuned so that they do not lose the impact of pre-training.  The audio-visual model is trained with different alignment, fusion and location strategies. Alignment methods are compared for mask fusion with concatenation, and we compare addition and concatenation for mask fusion with upsampling. Four fusion locations are compared with upsampling based alignment and concatenative fusion corresponding to  \textit{input}, \textit{intermediate}, \textit{late} and \textit{mask} fusion.   

\subsection{Experimental Results}
Table \ref{tab:ns_eval_audioset} reports the performance of our models. The audio-visual noise suppression models that use object recognition and video classification features improves performance over the audio baseline on all metrics. By employing the proposed multi-task learning framework for visual supervision,  we find that enhancement performance improves further, and obtains the best PESQ score of 2.60.  We find that object and video-based features attain similar classification performance, with higher gains from object-based multi-task training. This is because most acoustic events labeled correspond to objects in the scene rather than actions. The multi-task framework improves over the audio-visual model in two cases- (a) when there are multiple distractor sources in the recording, and (b) when the distractor source has a distinct visual representation.  


Fig.~\ref{fig:pesq_fusion} highlights PESQ performance of our audio-visual models across alignment methods, fusion methods, and fusion locations. It is observed that attention and upsampling have comparable performance, as do addition and concatenation. This can be useful to reduce the number of model parameters since addition and upsampling involve no additional parameters.  Of the different locations for fusion, \textit{late fusion} and \textit{mask fusion}, where the features are integrated with the output of the recurrent processor or the complex mask respectively, perform the best. The observation that late fusion outperforms input fusion is  similar to those made for acoustic event detection with Audioset~\cite{li2021aed}. Our best model uses addition-based late fusion with deterministic upsampling. This implies that the visual information can be used to generate a corrective mask that mitigates errors within the audio-only mask, thereby improving speech enhancement performance.   

Table \ref{tab:dnsmos_eval} demonstrates that the proposed multi-task approach improves DNSMOS by 0.07 absolute overall. Further, on a AB test conducted using 20 test samples and 10 respondents, 48\% of the listeners preferred the multi-task model, while 46\% of listeners had no preference and 6\% of listeners preferred the audio only model.
\begin{table}[htp]
\centering
\caption{DNSMOS evaluation of noise suppression models}
\begin{tabular}{|l|r|r|r|}
\hline
\textbf{Model}     & \multicolumn{1}{|l|}{\textbf{OVL}} & \multicolumn{1}{|l|}{\textbf{SIG}} & \multicolumn{1}{|l|}{\textbf{BAK}} \\ \hline
Audio              & 3.22                              & 3.48                              & 4.12                              \\
Audio-visual (Obj) & 3.27                              & 3.52                              & 4.12                              \\
+ MTL              & 3.29                              & 3.53                              & 4.14      \\ \hline                       
\end{tabular}
\label{tab:dnsmos_eval}
\end{table}
\vspace{-4mm}


\subsection{Ablation Study}

\noindent \textbf{SNR}: Table \ref{tab:snr_analysis_pesq} describes the PESQ improvements obtained by using visual features across different SNRs. We note that the proposed visual features produce improvements across all SNRs. 

\begin{figure} [tbp]
    \centering

    \includegraphics[width=0.25\textwidth]{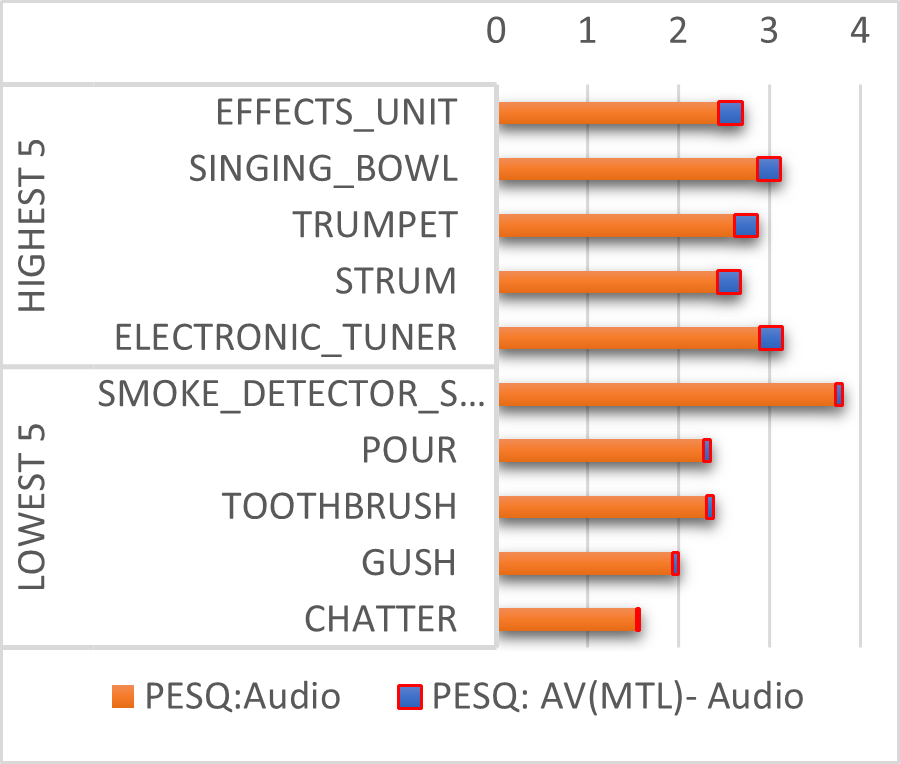}
        \caption{Plot showing noise label classes with 5 highest and 5 lowest changes in PESQ score between audio and audio-visual (MTL) models}
    \label{fig:pesq_comparison_noise_types}
\end{figure}

\noindent \textbf{Noise-type Analysis}: Audioset has human annotated acoustic event labels, which correspond to different types and sources of noise. The difference in PESQ scores between the baseline audio model and the proposed Audio-visual(MTL) model is computed, and the five classes with highest and lowest changes are shown in Fig.~\ref{fig:pesq_comparison_noise_types}. First, we observe that the PESQ score improves across \textit{all} noise types since the lowest change is positive. The improvements are highest over classes with distinct visual representations, i.e. ``strum'' sees higher gains compares to ``chatter''.

\noindent \textbf{Number of noise sources}: The number of such acoustic event labels from Audioset could also serve as a proxy for the number of distinct distractor sources. Fig.~\ref{fig:pesq_number_labels} plots the improvements in PESQ for different number of ground-truth Audioset labels in the distractor source. The data distribution (shown in green) demonstrates that a large proportion of the data has fewer than 5 labels. The audio-visual baseline model (yellow) seems to improve over the audio baseline for a single acoustic event, but degrades with more events. This is because of multiple reasons: (a) All the visual sources may not be on screen, or (b) Visual feature extractors may identify representations of other objects or actions that do not produce sounds. The difference between the proposed audio-visual (MTL) model and the audio-visual baseline (shown in blue) demonstrates the importance of visual supervision for multiple noise labels in the data. In conclusion, the proposed audio-visual (MTL) model improves over the audio baseline for single and multiple noise sources.


\section{Conclusion}
\label{sec:conclusion}
To the best of our knowledge, this paper is among the first work for egocentric audio-visual noise suppression. We consider the setting where the target speaker is off-screen with on-screen video comprising information regarding noise. We have introduced methods to  efficiently and accurately use the visual information. Specifically, we have investigated how to extract useful visual features and how to fuse them with the audio information.


 We evaluated different approaches for multimodal alignment and fusion, with additive mask fusion emerging as the best performing solution. To generate discriminative visual representations, we proposed a multi-task training framework that jointly optimizes audio-visual noise suppression and visual acoustic event detection. This multi-task training approach is shown to outperform the methods which directly uses object classification or action classification features as the input of the visual branch. This proposed framework resulted in 0.16 absolute PESQ improvement over a strong audio baseline. The proposed multi-task approach obtains improvements across all metrics, different SNR conditions and noise types; and also improves performance on audio with multiple distractor sources. 







\vfill\pagebreak

\bibliographystyle{IEEEbib}
\bibliography{refs.bib}
\label{sec:refs}

\end{document}